\documentclass[letterpaper,twocolumn,prb,aps,showpacs,superscriptaddress,
floatfix ]{revtex4-1}
\usepackage[latin1]{inputenc}
\usepackage{bm}
\usepackage[usenames]{color}
\usepackage{multirow}
\usepackage{amssymb}
\usepackage{amsbsy}
\usepackage{amsmath}
\usepackage{stmaryrd}
\usepackage{graphicx}
\usepackage{epsfig}
\usepackage{placeins}
\usepackage[normalem]{ulem}
\usepackage{bbold}
\usepackage{braket}
\usepackage{blindtext}
\usepackage[colorlinks,linkcolor=blue,citecolor=blue,urlcolor=blue]{hyperref}

\usepackage{scalerel}
\usepackage{tikz}
\usetikzlibrary{calc}
\usetikzlibrary{patterns}
\usetikzlibrary{svg.path}
\definecolor{orcidlogocol}{HTML}{A6CE39}
\tikzset{
  orcidlogo/.pic={
    \fill[orcidlogocol] svg{M256,128c0,70.7-57.3,128-128,128C57.3,256,0,198.7,0,128C0,57.3,57.3,0,128,0C198.7,0,256,57.3,256,128z};
    \fill[white] svg{M86.3,186.2H70.9V79.1h15.4v48.4V186.2z}
                 svg{M108.9,79.1h41.6c39.6,0,57,28.3,57,53.6c0,27.5-21.5,53.6-56.8,53.6h-41.8V79.1z M124.3,172.4h24.5c34.9,0,42.9-26.5,42.9-39.7c0-21.5-13.7-39.7-43.7-39.7h-23.7V172.4z}
                 svg{M88.7,56.8c0,5.5-4.5,10.1-10.1,10.1c-5.6,0-10.1-4.6-10.1-10.1c0-5.6,4.5-10.1,10.1-10.1C84.2,46.7,88.7,51.3,88.7,56.8z};
  }
}

\newcommand\orcid[1]{\href{https://orcid.org/#1}{\mbox{\scalerel*{
\begin{tikzpicture}[yscale=-1,transform shape]
\pic{orcidlogo};
\end{tikzpicture}
}{|}}}}

\makeatletter

\begin{document}

\title{Quantum versus Classical Dynamics in Spin Models: Chains, Ladders, 
and Square Lattices}

\author{Dennis Schubert}
\email{dennis.schubert@uos.de}
\affiliation{Department of Physics, University of Osnabr\"uck, D-49069 
Osnabr\"uck, Germany}

\author{Jonas Richter \orcid{0000-0003-2184-5275}}
\email{j.richter@ucl.ac.uk}
\affiliation{Department of Physics and Astronomy, University College London, 
Gower Street, London WC1E 6BT, UK}

\author{Fengping Jin}
\affiliation{Institute for Advanced Simulation, J\"ulich Supercomputing Centre,
Forschungszetrum J\"ulich, D-52425 J\"ulich, Germany }

\author{Kristel Michielsen}
\affiliation{Institute for Advanced Simulation, J\"ulich Supercomputing Centre,
Forschungszetrum J\"ulich, D-52425 J\"ulich, Germany }

\author{Hans De Raedt}
\affiliation{Zernike Institute for Advanced Materials University of Groningen, 
NL-9747 AG Groningen, 
Netherlands}

\author{Robin Steinigeweg \orcid{0000-0003-0608-0884}}
\email{rsteinig@uos.de}
\affiliation{Department of Physics, University of Osnabr\"uck, D-49069 
Osnabr\"uck, Germany}

\date{\today}


\begin{abstract}

We present a comprehensive comparison of spin and energy dynamics in quantum 
and 
classical spin models on different geometries, ranging from 
one-dimensional chains, over quasi-one-dimensional ladders, to two-dimensional 
square lattices. Focusing on dynamics at formally infinite temperature, we 
particularly consider the autocorrelation functions of local densities, where 
the time evolution is governed either by the linear Schr\"odinger equation in 
the quantum case, or the nonlinear Hamiltonian equations of motion in the case 
of classical mechanics. While, in full generality, a quantitative agreement 
between quantum and classical dynamics can therefore not be expected, our 
large-scale numerical results for spin-$1/2$ systems with up to $N = 36$ 
lattice sites in fact defy this expectation. Specifically, we observe a 
remarkably good agreement for all geometries, which is best for 
the nonintegrable quantum models in quasi-one or two dimensions, but still 
satisfactory in the case of integrable chains, at least if transport 
properties are not dominated by the extensive number of conservation laws.    
Our findings indicate that classical or semi-classical simulations provide a 
meaningful strategy to analyze the dynamics of quantum many-body models, even 
in cases where the spin quantum number $S = 1/2$ is small and far away from the 
classical limit $S \to \infty$.

\end{abstract}

\maketitle


\section{Introduction}
\label{Sec::Introduction}

Understanding the properties of quantum many-body systems out of equilibrium is 
a notoriously difficult task with relevance to various 
areas of 
modern physics, ranging from fundamental aspects of statistical mechanics  
\cite{polkovnikov2011, eisert2015} to more applied issues in material science 
and quantum information technology. Quantum spin systems are of particular 
importance in this context, since they describe the 
magnetism 
of certain compounds in nature \cite{mikeska2004}, can be 
realized in new experimental 
platforms \cite{bloch2008, langen2015}, or can be simulated on already 
available 
or future quantum computers \cite{Smith_2019, richter2021}. 

From a theoretical point of 
view, quantum spin systems routinely serve as test beds to study concepts such 
as the eigenstate thermalization hypothesis \cite{deutsch1991, srednicki1994, 
rigol2008, 
dalessio2016, borgonovi2016} or the phenomenon of many-body localization 
\cite{nandkishore2015, abanin2019}. Moreover, in the case of one-dimensional 
chain geometries, the integrability of certain spin models, accompanied by the 
existence of an extensive set of (quasi-)local conserved charges 
\cite{prosen2011, prosen2013, 
ilievski2016}, paves the way to obtain analytical insights, e.g., 
regarding their transport and relaxation behavior in the thermodynamic limit 
\cite{bertini2021, 
bulchandani2021, Castro_Alvaredo_2016, Bertini_2016}. At the same time, the 
development of sophisticated numerical techniques \cite{schollwoeck2005, 
schollwoeck2011, 
prelovsek2013} has significantly advanced our understanding of 
out-of-equilibrium processes in quantum spin models. Yet, most of these 
methods 
are best suited for \text{(quasi-)}one-dimensional situations, while the 
numerical 
treatment of spin systems in higher dimensions continues to be a hard 
task due to the exponentially growing Hilbert space and the fast build-up 
of entanglement \cite{czarnik2019, 
gan2020, verdel2021,leviatan2017, richter2020-2, denicola2021}.

As opposed to quantum systems, the phase space of classical systems grows only 
linearly with the number of constituents, such that simulations of systems with 
several thousands of lattice sites pose no problem and higher dimensions are 
feasible with today's machinery as well.
In fact, ranging back to the seminal work by Fermi, Pasta, Ulam, and Tsingou 
\cite{FPU2008}, numerical simulations of equilibration and thermalization in 
classical many-body systems have a long history \cite{windsor1967, lurie1974}. 
In particular, most relevant in the context of the present work, transport of  
spin and 
energy in classical spin models has been scrutinized 
extensively over the past decades \cite{mueller1988, gerling1989, gerling1990, 
 dealcantarabonfim1992,  dealcantarabonfim1993, boehm1993, Srivastava_1994, 
Constantoudis_1997, 
Oganesyan_2009, steinigeweg2012, Huber_2012, bagchi2013, Prosen_2013, 
Jen_i__2015, Das_2019_2, li2020, Glorioso_2021}.  
However, within the large body of literature on classical spin systems 
 \cite{mueller1988, gerling1989, gerling1990, 
 dealcantarabonfim1992,  dealcantarabonfim1993, boehm1993, Srivastava_1994, 
Constantoudis_1997, 
Oganesyan_2009, steinigeweg2012, Huber_2012, bagchi2013, Prosen_2013, 
Jen_i__2015, Das_2019_2, li2020, Glorioso_2021, lagendijk1977, deraedt1981,  
Steinigeweg2009, jin2013, Das_2018}, less attention has been devoted to a 
quantitative comparison of dynamics in classical and quantum spin models 
\cite{Gamayun_2019, richter2020,Elsayed2015,Starkov2018,Starkov2020}. Such a comparison is in the center of the 
present paper. 

On the one hand, in 
the case of quantum dynamics, the time evolution is governed by the linear 
Schr\"odinger equation and, for certain 
one-dimensional models, integrability can strongly impact their 
dynamics, leading to nondecaying currents and ballistic 
transport due to overlap with the extensively many conservation laws. On the 
other hand, classical spin systems evolve according 
to the nonlinear Hamiltonian equations of motion, and (except for some notable 
examples \cite{Das_2019, Prosen_2013}) even one-dimensional chains are 
nonintegrable and highly chaotic \cite{de_Wijn_2012}. While it seems likely that 
quantum and 
classical 
systems become more and more similar if the spin quantum number $S$ 
is successively increased from $S = 1/2, 1, \dots$ 
\cite{dupont2020, Richter_2019_3} towards the 
classical limit $S \to \infty$, it still is a non-trivial 
question whether 
and to which degree their dynamics agree with each other. While 
substantial differences most likely emerge at low temperatures 
$T$, a quantitative 
agreement between quantum and classical dynamics can, in full generality, not 
be expected at high temperatures either, especially when considering the 
most quantum case $S = 1/2$. In particular, integrability of 
certain $S = 1/2$ models reflects itself in their dynamics even at $T\to 
\infty$. Moreover, certain phenomena, such as the onset of many-body 
localization in strongly disordered quantum systems, have no classical 
counterpart such that an agreement between quantum and classical dynamics 
is unlikely in these cases \cite{richter2020,ren2020}.

\begin{figure}[t]
\centering
\includegraphics[width=.45\textwidth]{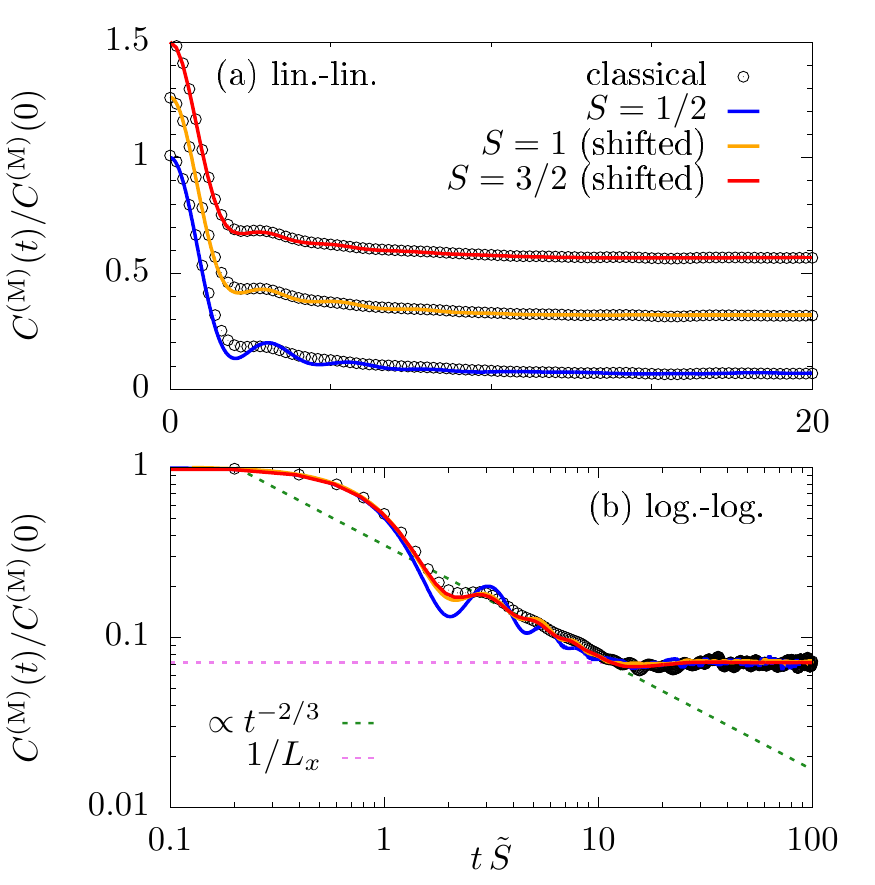}
\caption{(Color online) {\it Magnetization and 1D chain.} Decay of the 
equal-site correlation $C^\mathrm{(M)}(t)$ in different quantum cases $(S = 
1/2$, $1$, and $3/2$) and in the classical case ($S = \infty$), shown in a (a) 
lin.-lin.\ plot and (b) log.-log.\ plot. In all cases, we have length $L_x = 
14$ and anisotropy $\Delta = 1$. In (a), curves are shifted for better 
visibility. In (b), a power law $\propto t^{-2/3}$ and the expected long-time 
value $C(t \to \infty) = 1/L_x$ are indicated.}
\label{fig::2}
\end{figure}

In this paper, we explore the question of quantum versus classical dynamics 
in spin systems by analyzing time-dependent autocorrelation functions of local 
densities [as defined below in Eq.~\eqref{Eq::CorrelationGeneral}], which are 
intimately related to transport processes in these models and have been studied 
before, both in the classical and the quantum case \cite{mueller1988, 
gerling1989, gerling1990, bagchi2013, li2020, richter2020, wurtz2020}. Our 
main finding is exemplified in 
Fig.~\ref{fig::2}, which shows the temporal decay of infinite-temperature spin 
autocorrelation functions $C^{(\text{M})}(t)$ in isotropic Heisenberg chains 
with 
different quantum numbers $S = 1/2, 1, 3/2$ and $S = \infty$ (classical). As 
becomes apparent from Fig.~\ref{fig::2}~(a), quantum and classical dynamics 
agree 
very well with each other on short as well as long time scales, and for all 
values of $S$ shown here. While the agreement is slightly better for larger $S$, 
it is still convincing for $S = 1/2$, where the quantum chain is integrable 
whereas the classical model is not. Moreover, plotted in a double-logarithmic 
representation [Fig.~\ref{fig::2}~(b)], we find that the hydrodynamic power-law 
tail $C^{(\text{M})}(t) \propto t^{-\alpha}$ at intermediate times is well 
described 
by $\alpha \approx 2/3$, which suggests superdiffusive transport within the 
Kardar-Parisi-Zhang (KPZ) universality class \cite{bulchandani2021, 
bertini2021, dupont2020, Gopalakrishnan_2019, Das_2019, Ljubotina_2017} (for 
more details 
see Sec.\ \ref{Subsec::ResultsMagnetizationChain} below). 

The remarkable agreement of quantum and classical dynamics in Fig.~\ref{fig::2} 
provides the starting point for the further explorations in this paper. 
Specifically, while Fig.~\ref{fig::2} shows results for short chains with $L = 
14$ (which is already quite demanding for $S = 3/2$), we 
particularly focus on a more 
in-depth comparison between $S = 1/2$ and $S = \infty$ using large-scale 
numerical simulations of XXZ models on different lattice geometries, which 
range 
from one-dimensional (1D) chains, over quasi-1D two-leg ladders, to 
two-dimensional (2D) square lattices; see Fig.~\ref{fig::1}. Relying on an 
efficient typicality-based pure-state propagation \cite{heitmann2020, jin2021}, 
we treat spin-$1/2$ systems with up to $N = 36$ lattice sites and study the 
agreement of quantum and classical spin and energy dynamics depending on the 
exchange anisotropy of the XXZ model and the lattice geometry chosen. In doing 
so, we find a remarkably good agreement for all lattice geometries, which is 
best for nonintegrable quantum models in quasi-one or two dimensions, and (as 
already indicated in Fig.\ \ref{fig::2}) still convincing for integrable quantum 
chains, at least in cases where transport is not ballistic due 
to the extensive 
set of conservation laws.  

The rest of this paper is structured as follows. First, we introduce in Sec.\ 
\ref{Sec::ModelsObservables} the considered models and observables in the 
quantum case and discuss their classical counterparts as well. Here, we also 
comment on the diffusive decay of equal-site autocorrelations. Then, we 
describe in Sec.\ \ref{Sec::NumericalTechniques} the numerical techniques used 
by us, where we focus on the concept of dynamical quantum typicality. 
Eventually, we present our 
numerical results in Sec.\ \ref{Sec::Results} and compare classical and quantum 
dynamics of local magnetization and energy in different lattice geometries. We 
summarize and conclude in Sec.~\ref{Sec::Summary}.

\section{Models and Observables} 
\label{Sec::ModelsObservables}

\subsection{Models}\label{Subsec::Models}

In this paper, we consider the anisotropic Heisenberg model (XXZ model) on a 
rectangular lattice with periodic boundary conditions (PBC), consisting of $N = 
L_{x} \times L_{y}$ sites in total, 
where $L_{x}$ and $L_{y}$ are the lattice extension in $x$ and $y$ direction, 
respectively. The Hamiltonian is given by, 
\begin{equation} \label{Eq::GeneralHamiltonian}
\mathcal{H} = J \sum_{\langle {\bf r}, {\bf r}' \rangle} h_{{\bf r}, 
{\bf r}'} \, ,
\end{equation}
where the sum runs over all bonds $\langle {\bf r}, {\bf r}' \rangle$ of 
nearest-neighboring sites ${\bf r} = (i,j)$ and ${\bf r}' = (i',j')$. The 
antiferromagnetic exchange coupling constant $J > 0$ is set to $J = 1$ in the 
following. The local terms in Eq.\ (\ref{Eq::GeneralHamiltonian}) read
\begin{equation} \label{Eq::LocalTerm}
h_{{\bf r}, {\bf r}'} = S^{x}_{{\bf r}} S^{x}_{{\bf r}'} + S^{y}_{{\bf r}} 
S^{y}_{{\bf r}'} + 
\Delta S^{z}_{{\bf r}} S^{z}_{{\bf r}'} \, ,
\end{equation}
where $\Delta$ parametrizes the anisotropy in $z$ direction and the components 
$S^{\mu}_{{\bf r}} , \mu \in \lbrace x,y,z\rbrace$ are spin-$S$ operators at 
site 
${\bf r}$, which fulfill the usual commutator relations ($\hbar = 1$)
\begin{equation}\label{Eq::CommutatorRelation}
[ S^{\mu}_{{\bf r}}, S^{\nu}_{{\bf r}'} ] = \imath \, \delta_{{\bf r}{\bf r}'} 
\,  
\epsilon_{\mu\nu\lambda} \, S^{\lambda}_{{\bf r}} \, ,
\end{equation}
where $\delta_{{\bf r}{\bf r}'}$ is the Kronecker-Delta symbol and 
$\epsilon_{\mu \nu 
\lambda}$ is the antisymmetric Levi-Civita tensor. For the specific case of $S = 
1/2$, these components can be expressed in terms of Pauli matrices, 
$S^{\mu}_{{\bf 
r}} 
= \sigma^{\mu}_{{\bf r}}/2$.

While total 
energy is naturally conserved, i.e., $[{\cal H},{\cal H}] = 0$, ${\cal H}$ is 
also invariant under 
rotation about the $z$ axis, i.e., the total magnetization in this direction is 
preserved for all $\Delta$,
\begin{equation} \label{Eq::MagnetizationConservation}
[ S^z, {\cal H} ] = 0 \, , \quad S^z = \sum_{{\bf r}} S^z_{{\bf r}} \, 
.
\end{equation}
In this paper, we consider the spin- and 
energy-transport properties of ${\cal 
H}$ depending on the lattice geometry, the value of $\Delta$, and the model 
being quantum or classical.
In particular, we study three special cases of the
$L_{x} \times L_{y}$ lattice: (i) $L_{y} = 1$, i.e., a one-dimensional 
chain; (ii) $L_{y} = 2$, i.e., a quasi-1D two-leg ladder; and
(iii) $L_{y} = L_{x}$, i.e., a two-dimensional square lattice; see 
the sketch in Fig.\ \ref{fig::1}. Concerning integrability, it is well known that 
the spin-$1/2$ chain is integrable in terms of the Bethe ansatz independent of 
the value of $\Delta$
\cite{Bethe1931,Levkovich2016}, while integrability 
is 
broken for models with either $S > 1/2$ or $D > 1$. This 
integrability will play a crucial role for our comparison of quantum and 
classical dynamics below. Specifically, it is well known that energy 
transport is purely ballistic in the integrable 
quantum chain, which will be in stark contrast to the 
dynamics of the chaotic classical chain. At the same time, integrability as 
such not necessarily rules out that quantum and classical transport properties 
can agree with each other. For instance, as demonstrated below, both the 
quantum and classical chain show diffusive spin transport for $\Delta > 
1$.

\subsection{Observables} \label{Subsec::Observables}

\begin{figure}[t]
\centering
\includegraphics[width=.45\textwidth]{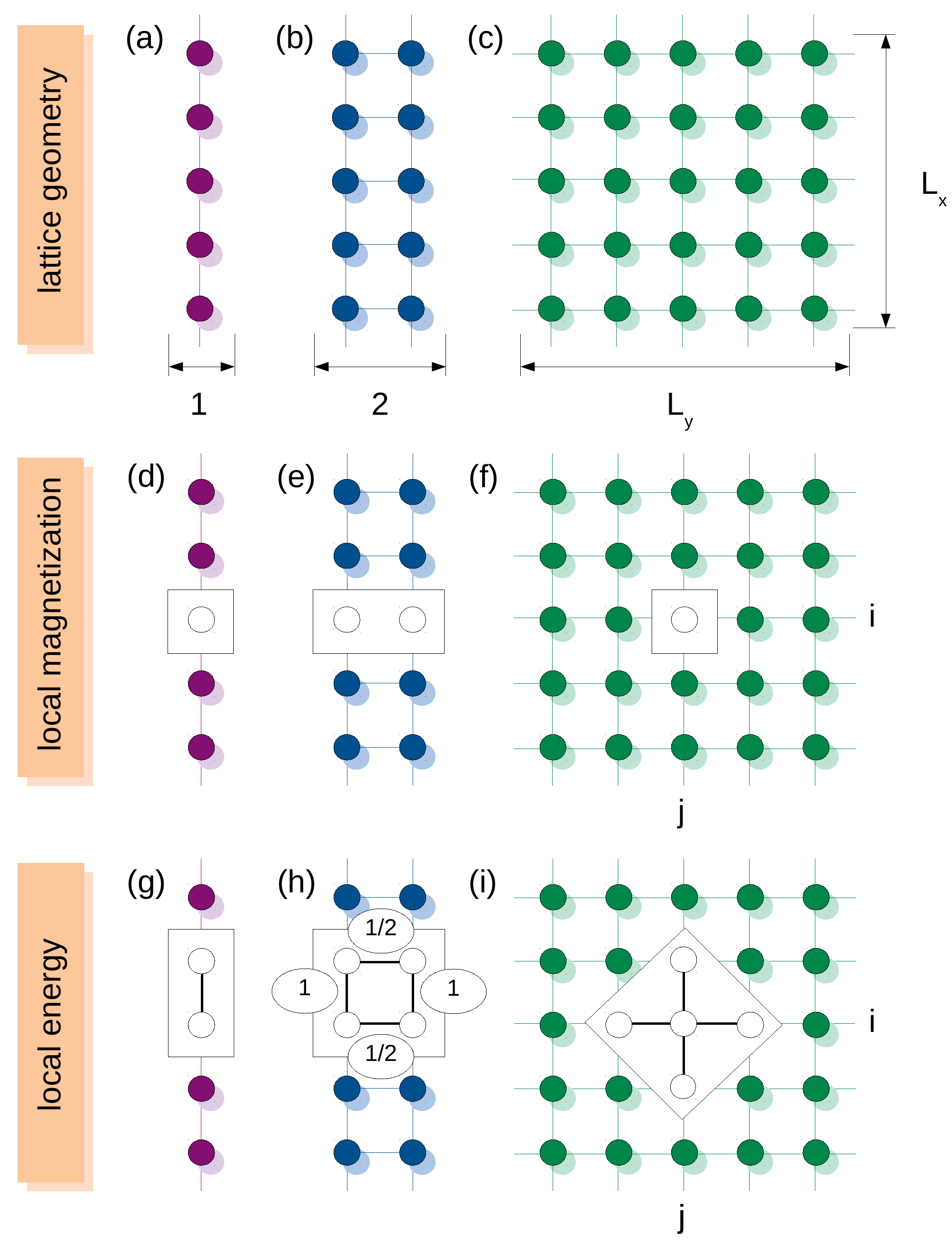}
\caption{(Color online) Overview over the different models and observables 
considered. Top row: (a) One-dimensional (1D) chain, (b) quasi-1D two-leg 
ladder, and (c) two-dimensional (2D) square lattice. Middle and bottom row: 
Corresponding local (d)-(f) magnetizations and (g)-(i) energies.}
\label{fig::1}
\end{figure}

As one of the simplest quantities, we focus on the dynamics of local densities 
$\rho_{{\bf r}}$, 
which can be either magnetization or energy, as defined below in detail. More 
precisely, we consider the time-dependent density-density correlation function, 
\begin{equation} \label{Eq::CorrelationGeneral}
C_{{\bf r}, {\bf r}'}(t) = \langle \rho_{{\bf r}}(t) \rho_{{\bf r}'} \rangle \, 
,
\end{equation}
where $\langle \bullet \rangle = \mathrm{tr}[\exp (-\beta {\cal H}) \bullet] / 
{\cal Z}$ with ${\cal Z} = \mathrm{tr}[\exp (-\beta {\cal H})]$ is a canonical 
expectation value at inverse temperature $\beta = 1/T$ ($k_\mathrm{B} = 1$), 
and the time argument of an operator has to be understood w.r.t.\ the 
Heisenberg picture, $\rho_{{\bf r}}(t) = \exp(\imath {\cal H} t) 
\, \rho_{{\bf 
r}} \, 
\exp(-\imath {\cal H} t)$.

In the following, we discuss the equal-site autocorrelation function, i.e., 
${\bf r} = {\bf r}'$ in Eq.~\eqref{Eq::CorrelationGeneral}. 
Due to our choice of PBC, the autocorrelation function does not depend on the 
specific site ${\bf r} = (i,j)$ and we can concisely write $C(t) = C_{{\bf r}, 
{\bf r}}(t)$.
Moreover, we here focus on the limit of high temperatures 
$\beta \to 0$ for which $\exp(-\beta {\cal H})/{\cal Z} \to \mathbb{1}/{\cal 
D}$, such that $C(t)$ is given by,
\begin{equation} \label{Eq::CorrelationInfiniteTemp}
C(t) = \frac{\mathrm{tr}[\rho_{{\bf r}}(t) \rho_{{\bf r}}]}{\cal D} \, ,
\end{equation}
where ${\cal D} = (2S+1)^{N}$ is the Hilbert-space dimension, e.g., 
${\cal D} = 2^L$ for $S = 1/2$. Note that for our numerical results, we always 
consider the dynamics in the full Hilbert space, i.e., we average over all 
sectors of fixed $S^z$. 

Next, we define the local densities $\rho_{{\bf r}}$ and start with the case of 
magnetization. While such a definition is not unique and depends on the 
chosen unit cell, we use the natural definition,
\begin{equation} \label{Eq::SumRule}
\rho_{i,j}^\mathrm{(M)} =
\left\{\begin{array}{ll}
S_{i,1}^z\ ,             & \text{1D } (L_y = 1)\\
S_{i,1}^z + S_{i,2}^z\ ,  & \text{quasi-1D } (L_y = 2)\\
S_{i,j}^z\ ,              & \text{2D} (L_x = L_y)
\end{array}\right. \, ,
\end{equation}
see the sketch in Fig.\ \ref{fig::1}. In the case of energy, a natural definition 
is,
\begin{equation}\label{Eq::Erg1}
\rho_{i,j}^\mathrm{(E)} = J \, h_{(i,1), (i+1,1)}\ , 
\end{equation}
for a 1D chain, i.e., just a single bond, and, 
\begin{eqnarray}
\rho_{i,j}^\mathrm{(E)} &=& J \, [ h_{(i,1), (i+1,1)} + 
h_{(i,2), (i+1,2)} ] \nonumber \\
&+& \frac{J}{2} \, [ h_{(i,1), (i,2)} + h_{(i+1,1), (i+1,2)} ]\ , 
\end{eqnarray}
for a quasi 1D two-leg ladder, i.e., a plaquette consisting of one bond for 
each leg and two rungs. Note that the factor $1/2$ appears, since the sum over 
all local 
energies must be identical to the total energy. For the 2D square lattice, we 
define, 
\begin{eqnarray}
\rho_{i,j}^\mathrm{(E)} &=& \frac{J}{2} \, [ h_{(i-1,j), (i,j)} + 
h_{(i,j), (i+1,j)} ] \nonumber \\ 
&+& \frac{J}{2} \, [ h_{(i,j-1), (i,j)} + h_{(i,j), (i,j+1)} ] \, 
,\label{Eq::Erg3}
\end{eqnarray}
see the sketch in Fig.\ \ref{fig::1} again.

We note that for each local density defined above, the sum rule $C(t = 0)$ 
can be calculated analytically. For instance, in the case of local 
magnetization, we have for $S=1/2$,
\begin{equation}
C^\mathrm{(M)}(t = 0) =
\left\{\begin{array}{ll}
1/4\ , & \quad \text{1D } (L_y = 1)\\
1/2\ ,  & \quad \text{quasi-1D } (L_y = 2)\\
1/4\ ,  & \quad \text{2D } (L_x = L_y)
\end{array}\right. \, .
\end{equation}
Assuming that the system thermalizes at long times, 
this initial value also determines the long-time value (although there can be 
subtleties in some cases, see Sec.\ \ref{Subsec::ResultsEnergy}), 
\begin{equation} \label{Eq::Saturation}
C(t \to \infty) = \frac{C(t = 0)}{n} \, ,
\end{equation}
where $n$ is the total number of unit cells, i.e., $n = L_{x}$ in 1D or 
quasi-1D and $n = L_{x} \times L_{y}$ in 2D. Therefore, only in the 
thermodynamic limit $n \to \infty$, we can expect a full decay $C(t \to 
\infty) 
= 0$.

\subsection{Classical limit}

The quantum spin models discussed so far also have a classical counterpart, 
which results by taking the limit of both, Planck's constant $\hbar \to 0$ and 
spin quantum number  $S \to \infty$, under the constraint $\hbar 
\sqrt{S(S+1)} = \text{const.}$ In this limit, the commutator relations in 
Eq.\ (\ref{Eq::CommutatorRelation}) then turn into,
\begin{equation} \label{Eq::PoissonRelation}
\{ S^{\mu}_{{\bf r}}, S^{\nu}_{{\bf r}'} \} = \delta_{{\bf r}{\bf r}'} 
\epsilon_{\mu \nu \lambda} \, S^{\lambda}_{{\bf r}} \, ,
\end{equation}
where $\{ \bullet, \bullet \}$ denotes the Poisson bracket 
\cite{Arnold1978}, and the spin operators become real three-dimensional vectors 
$\mathbf{S}_{{\bf r}}$ of constant length, $| \mathbf{S}_{{\bf r}} | = 1$. In 
particular, all symmetries mentioned before carry over to the classical case. 
The relations in Eq.\ (\ref{Eq::PoissonRelation}) lead to the Hamiltonian 
equations of motion, which read,
\begin{equation} \label{Eq::ClassicalHamiltonEquation}
\frac{\mathrm{d}}{\mathrm{d}t} \mathbf{S}_{{\bf r}} = 
\frac{\partial \mathcal{H}}{\partial \mathbf{S}_{{\bf r}}} 
\times \mathbf{S}_{{\bf r}}\ , 
\end{equation}
and describe the precession of a spin around a local magnetic field 
resulting from the interaction with the neighboring spins. The equations 
(\ref{Eq::ClassicalHamiltonEquation}) form a set of coupled differential 
equations, which is non-integrable by means of the Liouville-Arnold theorem 
\cite{Arnold1978, Steinigeweg2009}. Therefore, they can be solved 
analytically only for a small number of special initial configurations, and 
solving them for non-trivial initial states requires numerical techniques.

The infinite-temperature density-density correlation in Eq.\ 
\eqref{Eq::CorrelationInfiniteTemp} 
can be obtained in the classical case by taking $\langle \bullet 
\rangle$ as an average over trajectories in phase space, 
\begin{equation} \label{Eq::Samples}
C(t) \approx \frac{1}{R} \sum_{r=1}^R \rho_{{\bf r}}(t) \rho_{{\bf r}}(0)\ , 
\end{equation}
where the initial configurations $\rho_{{\bf r}}(0)$ are drawn at 
random for each realization $r$, 
and $R \gg 1$ has to be chosen sufficiently large to reduce statistical 
fluctuations. For the values of 
$R$ chosen by us, see the discussion in Sec.\
\ref{Subsec::ClassicalAveraging}.

In this paper, our central goal is to compare classical and quantum dynamics. 
Thus, for a fair comparison, we have to take into account that the sum 
rule $C(t = 0)$ is different. For instance, in the case of local magnetization, 
the classical sum rule is,
\begin{equation}
C^\mathrm{(M)}(t = 0) =
\left\{\begin{array}{ll}
1/3\ ,  & \quad \text{1D } (L_y = 1)\\
2/3\ ,  & \quad \text{quasi-1D } (L_y = 2)\\
1/3\ ,  & \quad \text{2D } (L_x = L_y)
\end{array}\right.\ ,  
\end{equation}
and differs from the one in Eq.\ (\ref{Eq::SumRule}). Thus, we always  
consider the rescaled data $C(t)/C(0)$, cf.~Fig.~\ref{fig::2}. Moreover, we 
have to 
rescale the time entering the quantum simulations by a factor 
\cite{richter2020},
\begin{equation}
\tilde{S} = \sqrt{S(S+1)} \, , 
\end{equation}
in order to account for the different length of quantum and classical spins 
($\tilde{S} = 1$ in the classical case). However, for $S = 1/2$, this factor is 
$\tilde{S} = \sqrt{3/4} \approx 0.87$ 
and rather close to $1$.

\subsection{Diffusion}

In both, the classical and the quantum case, the time evolution of the 
autocorrelation function $C(t)$ follows from the underlying microscopic 
equations 
of motion, and naturally depends on the specific model and its parameters. 
Thus, a precise statement on the functional form of this time evolution 
requires to 
solve the given many-body problem analytically or numerically. Due to the 
conservation of total energy and magnetization, however, one generally expects 
that the dynamics of local densities acquire a hydrodynamic behavior at 
sufficiently long times. In particular, in a generic 
nonintegrable situation, one might expect the emergence of normal diffusive 
transport.

In the context of the autocorrelation function $C(t)$, the emergence of 
hydrodynamics reflects itself in terms of a  a power-law tail 
\cite{bertini2021}, 
\begin{equation} \label{Eq::PowerLaw}
C(t) \propto t^{-\alpha} \, ,
\end{equation}
where normal diffusive transport corresponds to $\alpha = D/2$, where $D$ is 
the lattice dimension, i.e., $\alpha = 
1/2$ in 1D or quasi-1D, and $\alpha = 1$ in 2D. In contrast to the case of 
normal diffusion, anomalous superdiffusion (cf.\ Fig.~\ref{fig::2}) and 
subdiffusion go along with an 
exponent $\alpha > D/2$ and $\alpha < D/2$, respectively, while ballistic 
transport is indicated by $\alpha = D$. 

Clearly, such a hydrodynamic power-law decay can only set in for times $t > 
\tau$ after some 
mean-free time $\tau$. Moreover, due to the saturation at a value $C(t \to 
\infty) > 0$ in any finite system, diffusion must break down for long times. 
Thus, in our numerical simulations below, the power-law decay in Eq.\ 
(\ref{Eq::PowerLaw}) can only be expected to appear in an intermediate time 
window, as already demonstrated in Fig.\ \ref{fig::2} above. 

While the analysis of the particular type of transport for a given model and 
lattice geometry is not the main aspect of this paper, it naturally arises 
while comparing the spin and energy dynamics of quantum and classical systems 
in Sec.\ \ref{Sec::Results}. 

\section{Numerical techniques} \label{Sec::NumericalTechniques}

Next, we discuss the methods used in our numerical simulations, both for the 
quantum and the classical case. In the former, we particularly employ the 
concept of dynamical quantum typicality (DQT) which gives access 
to autocorrelation functions for comparatively large system sizes beyond the 
range of full exact diagonalizaton.

\subsection{Dynamical quantum typicality} \label{Subsec::DQT}

DQT essentially relies on the fact that even a single pure state $\ket{\psi}$ 
can imitate the full statistical ensemble. More precisely, the 
pure-state expectation value of an observable is typically close to the one in 
the statistical ensemble 
\cite{Lloyd2013, Gemmer2009, Goldstein2006, Reimann2007}.
This fact can be utilized to calculate the time 
dependence of correlation functions, e.g., the one of the density-density 
correlator in Eq.\ (\ref{Eq::CorrelationGeneral}), by replacing
the trace by a scalar product between two auxiliary pure states 
$\ket{\varphi_{\beta}(t)}$ and $\ket{\Phi_{\beta}(t)}$ \cite{Iitaka2004, 
Elsayed2013, Steinigeweg2014}, 
\begin{equation} \label{Eq::DQTApproximation}
C(t) = \frac{\langle \varphi_{\beta}(t) | \rho_{{\bf r}} 
| \Phi_{\beta}(t) \rangle}{\langle \varphi_{\beta}(0) | 
\varphi_{\beta}(0) \rangle} + \varepsilon (| \psi \rangle) \, , 
\end{equation}
where the two auxiliary pure states are given by,
\begin{eqnarray}
| \varphi_{\beta}(t) \rangle &=& e^{-\imath \mathcal{H} t}
e^{-\beta \mathcal{H}/2} \, | \psi \rangle \, , \label{State1}\\
| \Phi_{\beta}(t) \rangle &=& e^{-i\mathcal{H}t} \, \rho_{{\bf r}} \,
e^{-\beta \mathcal{H}/2} \, | \psi \rangle \, , \label{State2}
\end{eqnarray}
involving the reference pure state,
\begin{equation} \label{Eq::ReferenceState}
| \psi \rangle = \sum_{k = 1}^{\cal D} (a_{k} + \imath b_{k} ) | k \rangle \, .
\end{equation}
This reference pure state is drawn at random from the full Hilbert space
according to the unitary invariant Haar measure \cite{Bartsch2009}. In 
practice, for any given orthogonal basis $| k\rangle$, the coefficients 
$a_{k}$ and $b_{k}$ are drawn randomly from a Gaussian probability 
distribution with zero mean.

While the statistical error $\varepsilon ( | \psi \rangle )$ in Eq.\
(\ref{Eq::DQTApproximation}) depends on the specific realization of the random
$| \psi \rangle$, the standard deviation of this statistical error can be 
bounded from above \cite{jin2021},
\begin{equation}
\sigma(\varepsilon) \leq b \propto \frac{1}{\sqrt{{\cal D}_{\mathrm{eff.}}}} \, 
,
\end{equation}
where ${\cal D}_{\mathrm{eff.}} = \mathrm{tr} \{ \exp [-\beta (\mathcal{H} - 
E_{0} ) ] \}$
denotes an effective dimension and $E_{0}$ is the ground-state 
energy of ${\cal H}$. Thus, at high temperatures $\beta \to 0$, ${\cal 
D}_{\mathrm{eff.}} \to {\cal D} = (2S+1)^N$ and $\sigma(\varepsilon$) 
is negligibly small for the finite but large system sizes we are 
interested in. In turn, the typicality-based approximation in Eq.\ 
(\ref{Eq::DQTApproximation}) is very accurate even for a single $| \psi 
\rangle$, and no averaging is required.

In the high-temperature limit $\beta \to 0$, the correlation function $C(t)$ 
can also be approximated on the basis of just one auxiliary pure state
\cite{Richter2019},
\begin{equation} \label{Eq::OneAuxiliaryState}
| \psi'(t) \rangle = e^{-\imath {\cal H} t} \, | \psi'(0) \rangle \, , \quad | 
\psi'(0) \rangle = \frac{\sqrt{\rho_{{\bf r}} + c} \, | \psi 
\rangle}{\sqrt{\langle 
\psi | \psi \rangle}} \, ,
\end{equation}
where $| \psi \rangle$ is again the reference pure state in 
Eq.\ (\ref{Eq::ReferenceState}) and the constant $c$ is chosen in such a 
way that $\rho_{{\bf r}} + c$ has non-negative eigenvalues. Then, the 
correlation 
function can be rewritten as a standard expectation value 
\cite{richter2020, richter2021, Chiaracane2021},
\begin{equation} \label{Eq::ExpectationValue}
C(t) = \langle \psi' (t) | \rho_{{\bf r}} | \psi' (t) \rangle + 
\varepsilon(| \psi \rangle) \, ,
\end{equation}
where we have employed $\mathrm{tr}[\rho_{{\bf r}}] = 0$. From a 
numerical point 
of view, Eq.~\eqref{Eq::ExpectationValue} is more efficient than Eq.\ 
\eqref{Eq::DQTApproximation} as only one state has to be evolved in time. It is 
crucial, however, that the square root of the operator in Eq.\ 
(\ref{Eq::OneAuxiliaryState}) can be carried out. In the case of local 
magnetization, this task is trivial, at least in the Ising basis. In the case of 
local energy, the task also is feasible and requires only a local basis 
transformation, involving a few lattice sites.

The central advantage of the typicality approximations in Eqs.\ 
\eqref{Eq::DQTApproximation} and \eqref{Eq::ExpectationValue} is the fact that 
the time dependence appears as a property of the pure states.
In particular, this time evolution can be obtained by an iterative 
forward propagation in real time,
\begin{equation}\label{TimeEvo}
| \psi'(t + \delta t) \rangle = e^{-\imath \mathcal{H} \delta t} \,
| \psi'(t) \rangle \, ,
\end{equation}
where $\delta t \ll J$ is a small discrete time step. Note that, even though 
not required 
for our purposes as we focus on $\beta = 0$, the action of $\exp 
(-\beta \mathcal{H}/2)$ in Eqs.\ \eqref{State1} and \eqref{State2} can 
be obtained by an analogous forward propagation in imaginary time 
\cite{hams2000}.

While various sophisticated methods exist to approximate the action of the 
matrix 
exponential in Eq.\ \eqref{TimeEvo}, the massively parallelized simulations on 
supercomputers used by us rely on both, 
Trotter decompositions and Chebyshev-polynomial expansions 
\cite{Dobrovitski2003, Weisse2006}. Since the matrix-vector multiplications 
required in these methods can be carried out efficiently w.r.t.\ 
memory, it is possible to treat systems as large as $N = 36$ spins, or even 
more \cite{Richter_2019_2}.

\subsection{Classical averaging}
\label{Subsec::ClassicalAveraging}

In the classical case, we solve the Hamiltonian equations of motion in Eq.\ 
(\ref{Eq::ClassicalHamiltonEquation}) numerically by means of a fourth-order 
Runge-Kutta scheme (RK4), with a small time step $\delta t$. In particular, 
$\delta t$ is chosen small enough such that the total energy and the total 
magnetization of ${\cal H}$ are conserved to very high accuracy during the time 
evolution. (For other algorithms, see Ref.\ \onlinecite{krech1981}.)

Since classical mechanics is not concerned with the 
exponential growth of the Hilbert space with system size $N$, much
larger systems can be accessed in this case. In fact, as the phase space 
increases only linearly with $N$, several thousands of 
sites or 
more pose no problem. While we indeed present result for such large 
systems, we also 
consider classical chains with fewer sites $N \leq 
36$ to ensure a fair comparison with the quantum case. 

Importantly, there is no analogue of typicality in classical mechanics. Hence, 
to obtain the correlation function $C(t)$, just a single random initial 
configuration is 
not sufficient and an average over many samples $R \gg 1$ is needed instead, 
see Eq.\ (\ref{Eq::Samples}). As a consequence, the computational cost is mainly 
set by $R$ and not so much by $N$. For instance, in our numerical simulations 
below, we will use as many samples as $R = {\cal O}(10^5)$, to ensure that 
the calculation of the correlation function goes along with small statistical 
errors. Note that the choice of a proper $R$ also depends on the considered 
time scale, i.e., a good signal-to-noise ratio at long times, where $C(t)$ has 
already decayed substantially, requires a larger value of~$R$. 

\section{Results}
\label{Sec::Results}

We turn to the discussion of our numerical results and start in 
Sec.\ \ref{Subsec::ResultsMagnetization} with the dynamics of local 
magnetization, where we particularly compare our classical and quantum results 
for the different cases of 1D chains (Sec.\ 
\ref{Subsec::ResultsMagnetizationChain}), quasi-1D two-leg ladders (Sec.\ 
\ref{Subsec::ResultsMagnetizationLadderSquare}), and 2D square lattices (Sec.\ 
\ref{Subsec::ResultsMagnetizationLadderSquare}). Corresponding 
results for the dynamics of local energy  are then presented in Sec.\
\ref{Subsec::ResultsEnergy}.

\subsection{Dynamics of local magnetization} 
\label{Subsec::ResultsMagnetization}

\subsubsection{1D chain}
\label{Subsec::ResultsMagnetizationChain}

We start with the dynamics of magnetization in a 1D chain. In Fig.\ 
\ref{fig::2} above, we have already presented results for the 
autocorrelation function $C^{(\text{M})}(t)$ at the isotropic point $\Delta = 
1$, 
where we have found that quantum dynamics for all quantum numbers $S = 
1/2,1,3/2$ 
agree remarkably well with the dynamics of the classical chain.    

\begin{figure}[b]
\centering
\includegraphics[width=.45\textwidth]{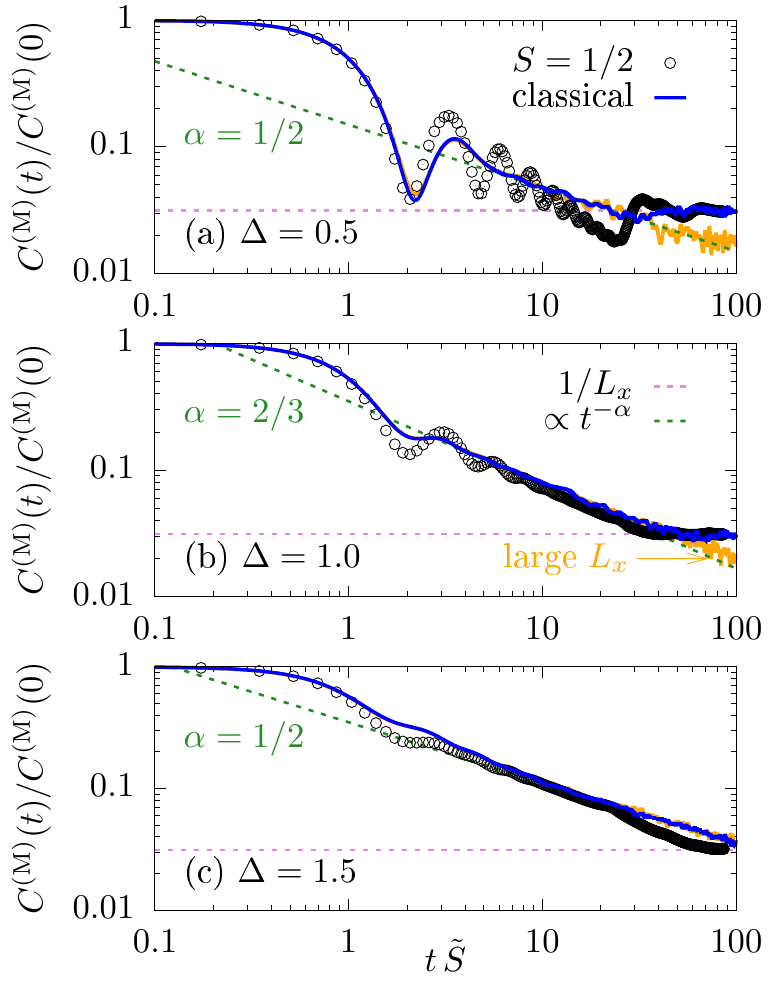}
\caption{(Color online) {\it Magnetization and 1D chain.} Decay of the 
equal-site correlation $C^\mathrm{(M)}(t)$ in a single quantum case $(S = 1/2$)
and in the classical case ($S = \infty$) for different anisotropies (a) $\Delta 
= 0.5$, (b) $\Delta = 1.0$, and (c) $\Delta = 1.5$, shown in a log.-log.\ plot.
In all cases, we have length $L_x = 32$ and indicate the expected long-time 
value $C(t \to \infty) = 1/L_x$ as well as power laws $\propto t^{-\alpha}$.
Classical data for a much larger $L_x = 1024$ are additionally depicted.}
\label{fig::3}
\end{figure}

\begin{figure}[t]
\centering
\includegraphics[width=.45\textwidth]{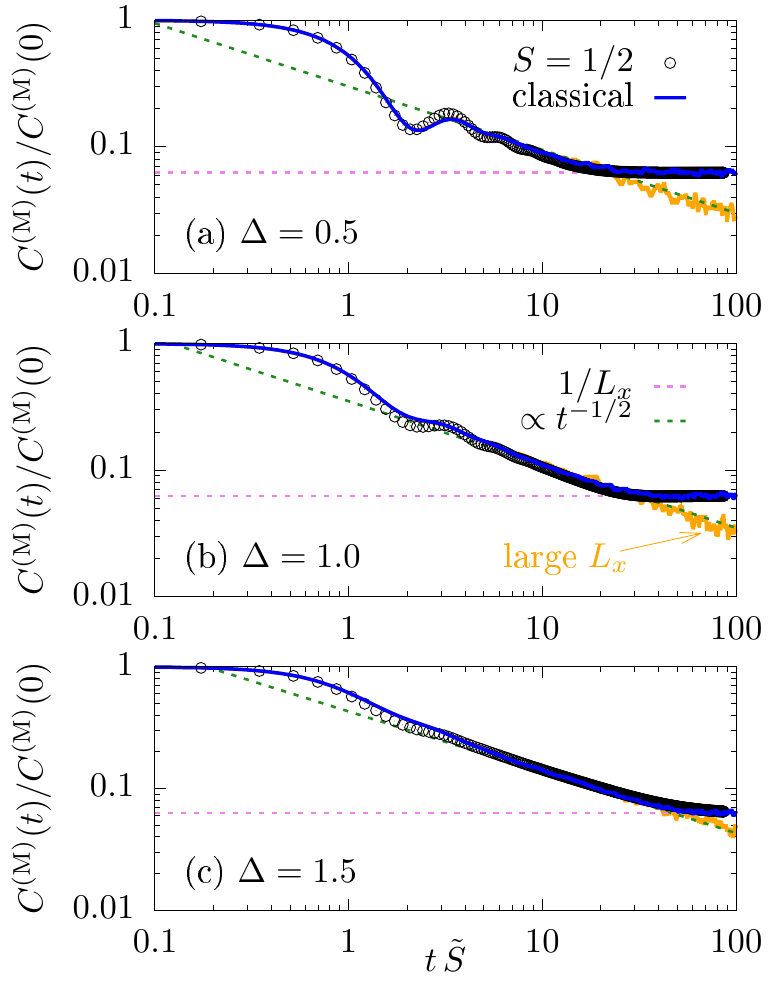}
\caption{(Color online) {\it Magnetization and quasi-1D two-leg ladder.}  
Relaxation of the equal-site correlation $C^\mathrm{(M)}(t)$ in the quantum 
case $(S = 1/2$) and in the classical case ($S = \infty$) for different 
anisotropies (a) $\Delta = 0.5$, (b) $\Delta = 1.0$, and (c) $\Delta = 1.5$, 
depicted in a log.-log.\ plot. In all cases, we have length $L_x = 16$ and 
indicate the expected long-time value $C(t \to \infty) = 1/L_x$ as well as a 
power law $\propto t^{-1/2}$. Classical data for a much larger $L_x = 512$ are
also shown.}
\label{fig::4}
\end{figure}

\begin{figure}[b]
\centering
\includegraphics[width=.45\textwidth]{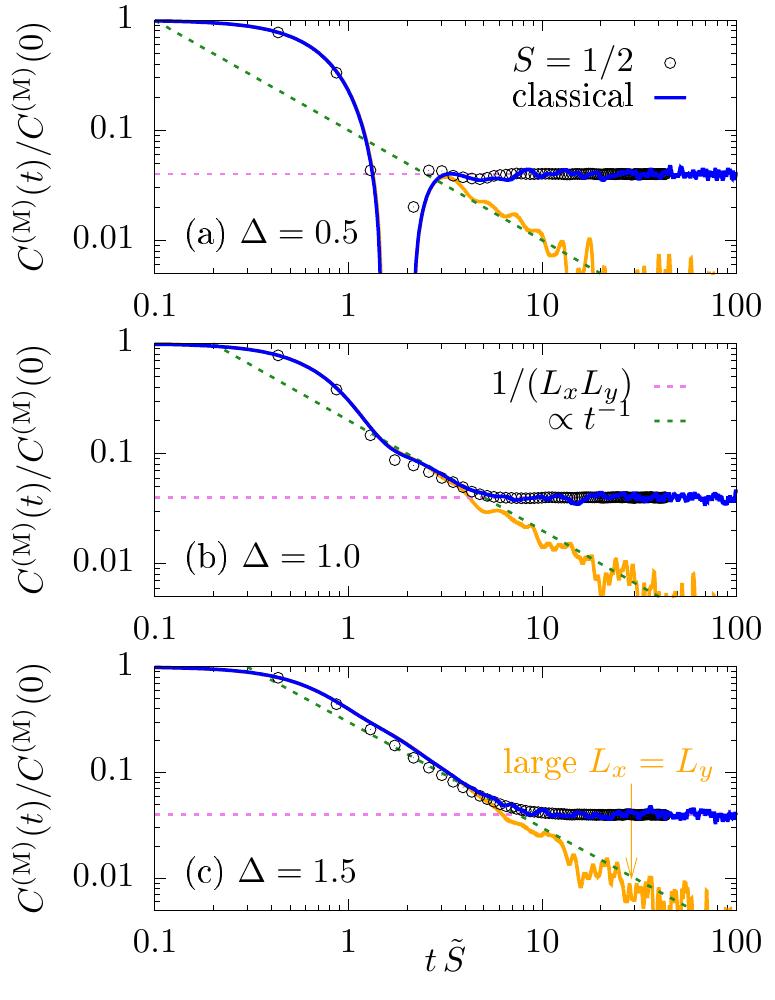}
\caption{(Color online) {\it Magnetization and 2D square lattice.} Time  
dependence of the equal-site correlation $C^\mathrm{(M)}(t)$ in the quantum 
case $(S = 1/2$) and in the classical case ($S = \infty$) for different 
anisotropies (a) $\Delta = 0.5$, (b) $\Delta = 1.0$, and (c) $\Delta = 1.5$, 
shown in a log.-log.\ plot. In all cases, we have edge length $L_x = L_y = 5$ 
and indicate the expected long-time value $C(t \to \infty) = 1/(L_x L_y)$ as 
well as a power law $\propto t^{-1}$. Classical data for a much larger $L_x = 
L_y = 32$ are also depicted.}
\label{fig::5}
\end{figure}

\begin{figure}[t]
\centering
\includegraphics[width=.45\textwidth]{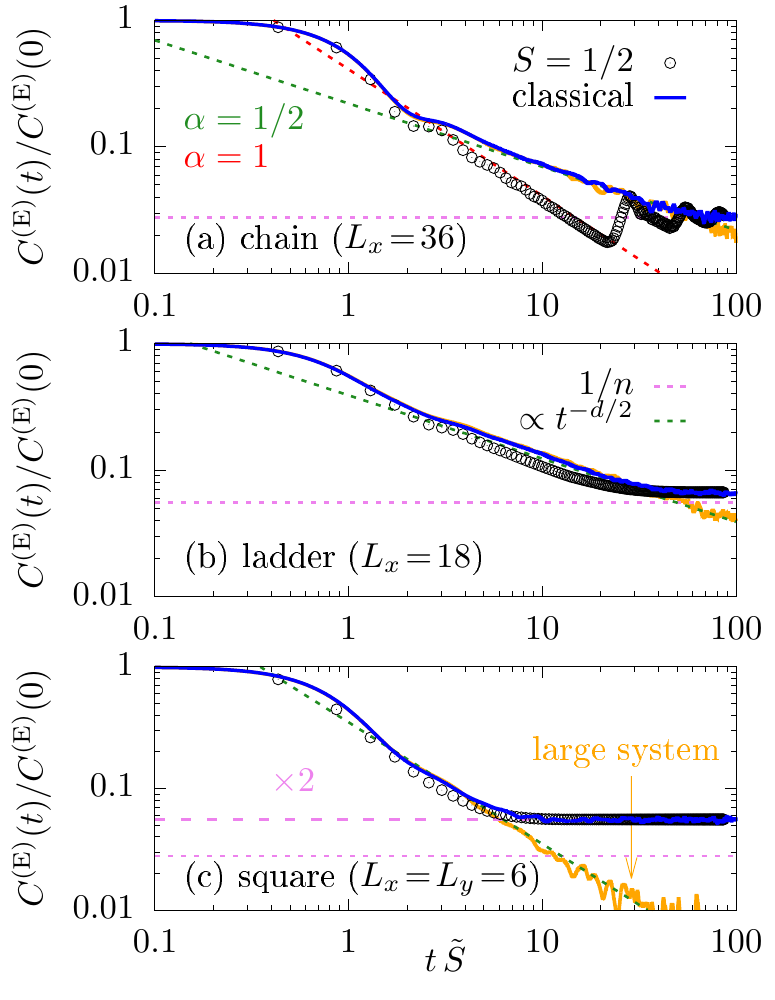}
\caption{(Color online) {\it Energy.} Relaxation of the equal-site correlation
$C^\mathrm{(E)}(t)$ in the quantum  case $(S = 1/2$) and in the classical case
($S = \infty$) for different lattice geometries, (a) 1D chain, (b) quasi-1D
two-leg ladder, and (c) 2D square lattice, depicted in a log.-log.\ plot. In 
all 
cases, we have anisotropy $\Delta = 1$ and indicate a power-law $\propto 
t^{-d/2}$. (Due to overlaps of local energies, the long-time value $C(t \to 
\infty)$ differs from $1/n$.) Classical data for a much larger $N = L_x \times 
L_y = 1024$ are also shown.}
\label{fig::6}
\end{figure}

Next, we discuss the role of the anisotropy $\Delta$, where we focus on the 
comparison between the most quantum case $S 
= 1/2$ and the classical case $S = \infty$. Thus, compared to Fig.\ 
\ref{fig::2}, we are able to access larger system 
sizes $L_x = 32 > 14$. In Fig.\ \ref{fig::3}, we summarize results 
for $C^\mathrm{(M)}(t)$ for anisotropies $\Delta = 0.5$, 
$1$, and $1.5$, in a double-logarithmic plot. For $\Delta = 1$ in Fig.\ 
\ref{fig::3}~(b), the situation is like the one in Fig.\ \ref{fig::2}~(b) 
discussed before. Due to the larger $L_x$, the long-time saturation value
becomes smaller and the power-law behavior persists on a longer time scale. 
Furthermore, when calculating classical data for a much larger $L_x = 
1024$, this range further increases. In particular, the data are still
consistent with an exponent $\alpha = 2/3$. On the one hand, in the case of the 
quantum chain, this superdiffusive behavior is by now well established at the 
isotropic point (see Ref.~\onlinecite{bulchandani2021} and references 
therein). On the other hand, in the case of the classical chain, the nature of 
spin transport at the isotropic point has been quite controversial 
\cite{mueller1988, gerling1989, gerling1990, 
 dealcantarabonfim1992,  dealcantarabonfim1993, boehm1993, Srivastava_1994}. 
While some recent works argue that the nonintegrability eventually causes the 
onset of normal diffusion with $\alpha = 1/2$ when going to sufficiently large 
systems and long time scales \cite{bagchi2013, li2020, dupont2020}, Ref.\ 
\onlinecite{De_Nardis_2020} provides compelling arguments that the power-law 
tail of $C^{(M)}(t)$ additionally acquires logarithmic corrections. Numerically, 
these scenarios are naturally very hard to distinguish.  

For the larger $\Delta = 1.5$ in Fig.\ \ref{fig::3} (c), we also observe a very 
good agreement between quantum and classical dynamics. Compared to $\Delta = 
1$, the main difference is a change of the exponent $\alpha$ from $2/3$ to 
$1/2$. Hence, this value indicates a diffusive decay, which is by now well 
known to occur the regime $\Delta > 1$, even in the case of the 
integrable quantum system \cite{bertini2021}. The results in 
Fig.\ \ref{fig::3}~(b) and (c) demonstrate that integrability of the quantum 
model as such not necessarily prevents that its dynamics are well approximated 
by a simulation of a classical system instead.  

For the 
smaller $\Delta = 0.5$ in Fig.\ \ref{fig::3} (a), we find a worse agreement 
between quantum and classical data, with oscillatory behavior for $S = 1/2$. 
While one might be tempted to conclude that the power-law decay of quantum 
and classical dynamics is similar at short times $t \lesssim 10$, such a 
conclusion is certainly not correct at longer times. On the one hand, as shown 
in 
Fig.\ \ref{fig::3} (a), classical dynamics for a long chain of length $L_x = 
1024$ is diffusive with $\alpha = 1/2$. On the other hand, quantum dynamics 
must be ballistic ($\alpha = 1$) in 
the thermodynamic limit, which has been proven rigorously using quasi-local 
conserved charges \cite{prosen2011, prosen2013, ilievski2016}.
Thus, in such cases, where the quantum dynamics is dominated by the extensive 
set of conservation laws, the remarkable correspondence between quantum and 
classical dynamics necessarily has to break down.  

\subsubsection{Quasi-1D two-leg ladder and 2D square lattice}
\label{Subsec::ResultsMagnetizationLadderSquare}

Next, we move from 1D chains to lattice geometries of higher dimension, i.e., 
quasi-1D two-leg ladders and 2D square lattices. By doing so, we break the 
integrability of the quantum system with $S = 1/2$. This non-integrable 
situation is certainly more generic and might be seen as a fair 
test bed for the comparison between the dynamics in models with $S = 1/2$ and 
$S = \infty$. As before, we focus on the decay of local magnetization and 
consider different values of the anisotropy $\Delta$.

For the quasi-1D two-leg ladder, we show in Fig.\ \ref{fig::4} the equal-site 
correlation $C^\mathrm{(M)}(t)$ for $\Delta = 0.5$, $1$, and $1.5$, where we 
fix the length of the ladder to $L_x = 16$. In contrast to the integrable case 
discussed 
before, we find a convincing 
agreement between quantum and classical relaxation for all three values of 
$\Delta$. In particular, the time dependence 
of $C^\mathrm{(M)}(t)$ at intermediate times turns out to be well described by 
a power law $t^{-\alpha}$ with the same diffusive exponent $\alpha = 1/2$ 
\cite{Richter_2019_2}.
For $L_x = 16$, this power-law behavior can be seen more clearly for larger 
$\Delta$ while, for classical systems with a much larger $L_x = 512$, it becomes 
even more pronounced. In view of non-integrability, the qualitative similarity 
of quantum 
and classical mechanics might not be too surprising. However, it is quite 
remarkable that the curves in Fig.\ \ref{fig::4} agree even on a quantitative 
level to high accuracy.

For the 2D square lattice, we summarize in Fig.\ \ref{fig::5} the decay of 
$C^\mathrm{(M)}(t)$ for the same values of $\Delta$ and a fixed edge length 
$L_x = L_y = 5$. The overall situation appears to be similar to the one for the 
quasi-1D two-leg ladder, e.g., the relaxation is well described by a power law 
$t^{-\alpha}$ with a diffusive exponent $\alpha$, which is $\alpha = 1$ in this 
2D case \cite{richter2021}. For $\Delta = 0.5$ in Fig.\ \ref{fig::5} (a), this 
power-law behavior 
cannot be seen at all for $L_x = L_y = 5$ due to finite-size effects, both for 
the quantum and the classical system.
However, when calculating  
classical data with a substantially larger $L_x = L_y = 32$, the 
diffusive decay eventually 
develops clearly also for $\Delta = 0.5$.

\subsection{Dynamics of local energy}
\label{Subsec::ResultsEnergy}

Finally, we turn to the dynamics of local energy. In this way, we want to 
ensure that the good agreement between quantum and classical dynamics is not 
restricted to the transport of local 
magnetization discussed above. For simplicity, 
let us focus on the isotropic point $\Delta = 1$ and study the impact of  
different lattice geometries.

In Fig.\ \ref{fig::6}, we show the time dependence of $C^\mathrm{(E)}(t)$ for a 
1D chain, a quasi-1D two-leg ladder, and a 2D square lattice, where we compare 
the dynamics of $S = 1/2$ and $S = \infty$ in finite systems. For the quasi-1D 
and 2D cases in Fig.\ \ref{fig::6} (b) and (c), we observe a very good
agreement between quantum and classical relaxation. However, for the 1D case in 
Fig.\ \ref{fig::6} (a), substantial differences can be clearly seen. In 
fact, these differences must occur as energy dynamics is ballistic ($\alpha 
= 1$) for $S = 1/2$ due to integrability \cite{zotos1997}, 
while the classical chain exhibits diffusive energy transport instead ($\alpha 
= 1/2$). Hence, Fig.\ \ref{fig::6}~(a), just like Fig.\ \ref{fig::3}~(a), 
constitutes a counterexample to our typical observation that the decay 
of quantum and classical density-density correlations agree qualitatively and 
quantitatively.

As a technical side remark, we note that the
energy-energy correlation functions saturate at a long-time value which 
disagrees with the naive prediction in Eq.\ \eqref{Eq::Saturation},  
\begin{equation}
C^\mathrm{(E)}(t \to \infty) \neq \frac{C^\mathrm{(E)}(0)}{n} \, . 
\end{equation}
This fact can be seen most clearly for the 2D square lattice in 
Fig.\ \ref{fig::6} (c). However, this observation should not be misunderstood 
as a breakdown of equipartition or thermalization.  
In fact, the prediction for the long-time value of $C(t)$ in Eq.\ 
\eqref{Eq::Saturation} generally is, 
\begin{equation}\label{Eq::Correct}
C(t \to \infty) = \frac{1}{n} \sum_{{\bf r}'} \langle \rho_{{\bf r}} \, 
\rho_{{\bf r}'} 
\rangle \, ,
\end{equation}
where the reference site ${\bf r}$ is fixed. We note that Eq.\ 
\eqref{Eq::Correct} is only identical to Eq.\ (\ref{Eq::Saturation}) if 
there is no overlap 
$\langle \rho_{{\bf r}} \, \rho_{{\bf r}'} \rangle$ between local densities at 
different sites. Such overlaps occur however naturally, given the definitions 
of 
the local energies in Eqs.\ \eqref{Eq::Erg1} - \eqref{Eq::Erg3}. For instance, 
for our choice of the local energy in 2D, the density $\rho_{i,j}^{(\text{E})}$ 
on site ${\bf r} = (i,j)$ shares a common bond with each of the four 
neighboring local energies $\rho_{i',j'}^{(\text{E})}$, with $i' = i \pm 1, j' 
= j \pm 1$, see Fig.\ \ref{fig::1}~(i). Hence, these bonds contribute to Eq.\ 
\eqref{Eq::Correct} 
and give rise to a correction by a factor of $2$, i.e.,
\begin{equation}
C^\mathrm{(E)}_\mathrm{2D}(t \to \infty) = 
\frac{C^\mathrm{(E)}_\mathrm{2D}(0)}{2 n} \, ,
\end{equation}
which is indicated in Fig.\ \ref{fig::6} (c) and coincides with the numerical 
simulation. Similar corrections apply to the long-time value of 
$C^{(\text{E})}(t)$ in chains and ladders as well, albeit they are less 
pronounced in these cases. 

\section{Summary} \label{Sec::Summary}

In this paper, we have addressed the question whether and to which 
degree the dynamics in spin systems with $S = 1/2$ and $S = \infty$ agree, 
focusing on the limit of high temperatures $T \to \infty$.
We have explored 
this question by studying XXZ models on different 
lattice geometries of finite size, ranging from 1D chains, over quasi-1D 
two-leg ladders, to 2D square lattices. In particular, we have analyzed 
the temporal decay of autocorrelation functions of local spin or 
energy densities, which are intimately related to transport properties in 
these models. In order to mitigate finite-size effects, we have relied on a 
combination of supercomputing and the typicality-based forward propagation of 
pure states, which has allowed us to treat quantum systems with up to $N = 36$ 
in 
total.    
As a main result, we have unveiled a remarkably good agreement between quantum 
and classical dynamics 
for all lattice geometries considered, which 
has been most pronounced for nonintegrable quantum systems in quasi-one or two 
dimensions.
Still, the agreement has turned out to be satisfactory also in the integrable 
quantum chain, at least in cases where the quantum dynamics is not ballistic 
due to the presence of additional conservation laws. Based on these findings, 
we conclude that classical or semi-classical/hybrid simulations might provide 
a meaningful strategy to investigate the quantum dynamics of strongly 
interacting quantum spin models, even if $S$ is small and far away from the 
classical limit.

While the numerical advantage of such classical simulations is obvious due to 
the substantially 
larger system sizes treatable, we have yet neither a rigorous argument for 
the good agreement observed nor an analytical estimate for the differences 
remaining. On the one hand, an approximate agreement between the quantum and 
classical versions of $C(t)$ might not be too surprising in cases where the 
quantum chain exhibits normal diffusive transport, as the emerging 
hydrodynamics on a coarse-grained level  should be effectively describable as a 
classical phenomenon. On the other hand, notwithstanding these arguments, the 
nice agreement between $S = 1/2$ and $S = \infty$ on a 
quantitative level, and on all time scales (even before the onset of 
hydrodynamics), remains remarkable to us. 

Our work raises a number of questions. First, it is not clear 
if a similar agreement between quantum and classical dynamics is expected for 
other observables beyond local densities or other out-of-equilibrium quantities 
beyond correlation functions. Secondly, another interesting direction of 
research is to clarify how far this agreement 
carries over to finite temperatures. Yet, it is clear that there should 
be some low-energy scale, where the specific excitations of a given quantum 
model become most relevant and likely cause large differences to the classical 
counterpart. Eventually, it would be interesting to compare the 
dynamics of quantum and classical models in the presence of disorder. While 
strongly disordered one-dimensional quantum models are believed to undergo a 
many-body localization transition, such a comparison would be particulary 
interesting in higher dimensions, where the fate of many-body localization is 
less clear.

\section*{Acknowledgments}

This work has been financially supported by the Deutsche Forschungsgemeinschaft 
(DFG), Grant No.\ 397067869 (STE 2243/3-1), within the DFG Research Unit FOR 
2692, Grant No.\ 355031190. J.\ R.\ has been funded by the European Research 
Council (ERC) under the European Union's Horizon 2020 research and innovation 
programme (Grant agreement No. 853368). Additionally, we gratefully acknowledge 
the 
computing time, granted by the ``JARA-HPC Vergabegremium'' and provided on the 
``JARA-HPC Partition'' part of the supercomputer ``JUWELS'' at 
Forschungszentrum J\"ulich.

\appendix

\begin{figure}[t]
\centering
\includegraphics[width=.45\textwidth]{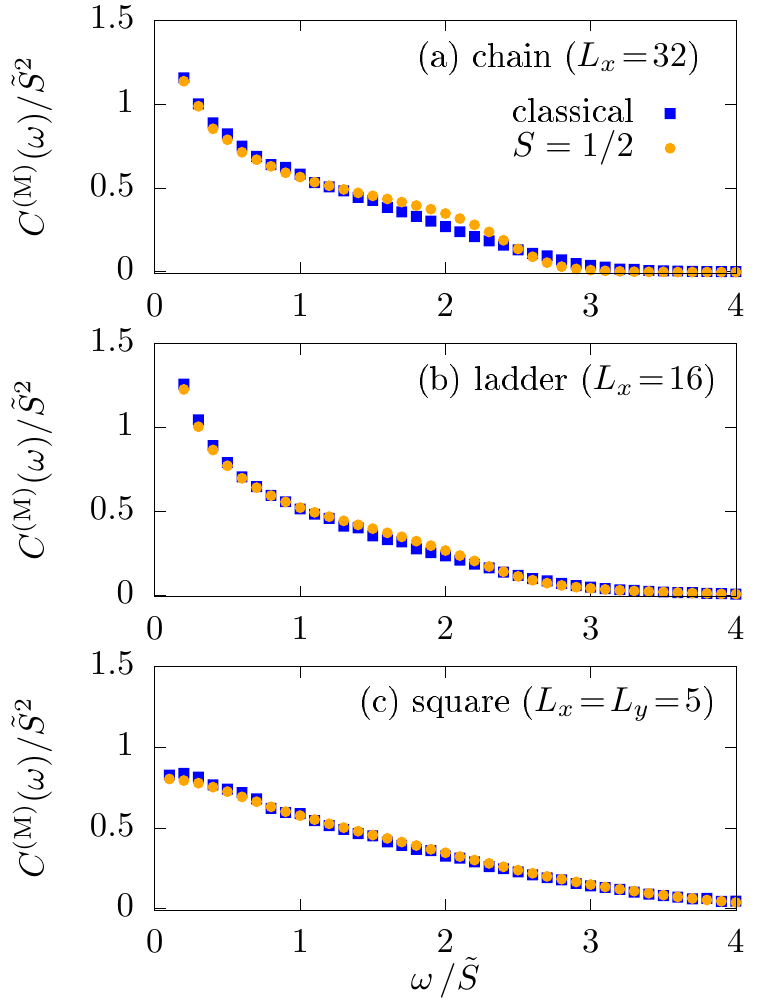}
\caption{(Color online) Magnetic spectral function $C^\mathrm{(M)}(\omega)$
in a single quantum case $(S = 1/2$) and in the classical case ($S = \infty$) 
for different lattice geometries, (a) 1D chain, (b) quasi-1D two-leg ladder, 
and (c) 2D square lattice, shown in a lin.-lin.\ plot. In all cases, we have 
anisotropy $\Delta = 1$. Data are obtained by Fourier transforming 
$C^\mathrm{(M)}(t)$ up to a cut-off time $t_\mathrm{max} \, \tilde{S}= 10 \pi$,
yielding a frequency resolution $\delta \omega / \tilde{S} = 0.1$.}
\label{fig::7}
\end{figure}

\section{Frequency space}

In the main text, we have focused on a comparison of quantum ($S=1/2$) and 
classical ($S=\infty$) mechanics in the time domain. This comparison could be 
done equally well in the frequency domain. Thus, in addition to the data for the 
correlation function $C(t)$ presented in Sec.\ \ref{Sec::Results}, we present 
here  data for the corresponding spectral function $C(\omega)$, which can be 
obtained from the Fourier transform
\begin{equation}
C(\omega) = \int_{-t_\mathrm{max}}^{t_\mathrm{max}} \mathrm{d}t \, e^{-\imath 
\omega t} \, C(t)
\,
\end{equation}
with a finite cut-off time $t_{\mathrm{max.}} < \infty$, yielding a frequency 
resolution $\delta \omega = \pi/t_\mathrm{max}$. In Fig.\ \ref{fig::7}, we 
exemplary depict the Fourier transform for the case of magnetization and 
anisotropy $\Delta = 1$. We do so for the 1D, quasi-1D, and 2D lattice 
geometry. Apparently, the agreement between quantum and classical mechanics 
is very good in the frequency domain as well.



%
\end{document}